# Phase relation investigation of U-La-O system under oxidizing conditions and observation of novel meta-stable and mixed-valent uranium phase- $Ln_3U_{11}O_{36}$ (Ln=La, Nd, Sm, Gd)

*Shafeeq Muhammed $^a$, Geeta Patkare $^a$, Rohan Phatak $^{a*}$*

$^a$ Fuel Chemistry Division, Bhabha Atomic Research Centre, Mumbai-400085, Maharashtra, India.

*corresponding author- email address: *raphatak@barc.gov.in, phatakrohan@gmail.com*

**Abstract:**

Total of twelve samples in the U-La-O system with the compositions $U_{1-y}La_yO_{2+x}$ (y=0.025, 0.05, 0.1, up to 0.3) were synthesized by gel combustion synthesis method followed by appropriate heat treatment in air atmosphere. Comprehensive experimental analysis using various techniques like X-ray diffraction, thermogravimetry and oxygen to uranium ratio (O/U) are used to establish the phase relation in $U_{1-y}La_yO_{2+x}$ system at 1173 K and 1523 K heated in air. A novel meta-stable phase with stoichiometry $La_3U_{11}O_{36}$ having mixed-valent uranium is reported for the first time in U-La-O system. The thermal property of this new phase is reported along with the updated phase relations at and above 1173 K temperature for the compositions $U_{1-y}La_yO_{2+x}$ (y ≤ 0.3). Further, formation and stability of this new $Ln_3U_{11}O_{36}$ phase was also investigated with smaller cations like $Nd^{3+}$, $Sm^{3+}$, $Gd^{3+}$ and the smallest $Y^{3+}$ cation.

**Introduction:**

$UO_2$ or other fissile/fertile isotope doped mixed oxides fuels (MOX) with fluorite structure are the most widely used nuclear fuels in BWR, PHWR, PFBR and many other proposed advanced reactor designs like Indian AHWR (Advanced Heavy Water Reactor) [1], CHTR (Compact High Temperature Reactor) [2] , etc. Fluorite type MOX fuels have the advantage of high temperature melting point as well as the structure has sufficient interstitial sites to accommodate fission products during the burn-up [3] and maintain the structural integrity in high radiation environment. The MOX fuels has been extensively studied for decades from all the aspects of the nuclear fuel cycle starting from synthesis, physical, chemical and thermal behavior, interaction with clad material, coolant and fission products, reprocessing of spent fuel and waste management.

Thermophysical properties of the fuel are one of the important aspects deciding the overall power of the specific reactor can achieve during operation. One of indirect but important property related to this is the oxidation behavior of the fuel during operational and accidental conditions.

Uranium exhibits multiple oxidation states and during oxidation, $UO_2$ (cubic) oxidizes to $U_3O_8$ (orthorhombic) which is disruptive in nature due to approximately 36% volume [4]. It is well known that aliovalent doping with $Ln^{3+}$ cation inhibits the formation of $U_3O_8$ phase, restricting it to fluorite structure. [5, 6]. Also, recent study on Ln-doped $UO_2$ samples by solution calorimetry indicate increased negative enthalpy of formation [7]. Apart from the interest for stabilizing fluorite phase, Ln- doped $UO_2$ matrix is exhaustively studied for its importance in simulating spent nuclear fuel (SNF) [8, 9] and advanced fuels with additives like gadolinium or erbium as burnable poison [10-12].

Many models have been proposed in the literature to explain the stability and reduced reactivity towards oxidation in Ln- doped $UO_2$ matrix. These models are briefly discussed by V. L. Vinograd et al. [13]. But the exact mechanism still remains unclear and detailed investigation of local structural, thermodynamic and oxidation kinetics is required to get a broader understanding of doped uranium oxides [14]. In our previous work on U-Nd-O system, we have shown that for all the compositions $U_{1-x}Nd_xO_{2\pm\delta}$ ($0.3 \leq x \leq 0.7$), the fluorite phase is stabilized even under oxidation conditions up to 1573 K temperatures. Within this range, oxygen to uranium ratio (O/U) varies from 2.5 and gradually increases to 3 with respective oxygen to metal ratio (O/M) varying from 2.21 to 1.95. Thus it is seen that fluorite phase is stable above $x \geq 0.3$ in $U_{1-x}Nd_xO_{2\pm\delta}$, below which O/M exceeds 2.25 limit with appearance of $(U,Nd)_3O_8$ phase. This observation is similar to fluorite phase limit of O/U ≤ 2.25 up to $U_4O_9$ in pure U-O system above which $U_3O_8$ phase exists. It is seen that charge compensation with $Ln^{3+}$ doping plays a major role of restricting the O/M values below 2.25 and thus stabilizing the fluorite structure. It was also observed that for freshly reduced samples in the series $U_{1-x}Nd_xO_{2-\delta}$ ($0.1 \leq x \leq 0.7$) having hypo-stoichiometry, the stability towards oxidation increases with increase in the Nd-concentration. This observation is indeed supported by many other experimental evidences [15, 16]. Prima facie, this indicates stabilization of the fluorite phase is due to charge compensation mechanisms in aliovalent substitution in $UO_2$. On the contrary, the stabilization of fluorite phase by isovalent substitution of Th in $(U_{1-y}Th_y)O_{2+x}$ system for compositions y > 0.2 [17] is surprising and shows that the stabilization of Fluorite phase and retardation towards oxidation are much more complex process.

In most of the previous work carried out on Ln-U-O systems, the samples were synthesized in the reducing atmosphere and later oxidized in air or synthesized at very high temperatures much above 1273 K. In our current study, we have tried to investigate how and at what compositions the initiation and stabilization of the fluorite phase takes place by synthesizing the samples in air without exposing it to inert or reducing atmosphere. The current synthesis approach and detailed XRD analysis help us to re-investigate the U-La-O system in air.

## 2. Experimental:

### 2.1 Synthesis:

Samples $U_{1-y}La_yO_{2+x}$ (y=0.025, 0.05,…0.3) were synthesized by gel- combustion method. $U_3O_8$ (Reactor grade) and $La_2O_3$ (AR grade) were used as a starting reagents. Both the reagents were preheated up to 1073 K and cooled to room temperature before using to remove any loose moisture adsorbed during storage. XRD patterns of the reagents were recorded to check the purity. Stock solutions of uranium and lanthanum were prepared by dissolving measure quantity of $U_3O_8$ and $La_2O_3$ in small quantity of concentrated nitric acid. Once the clear solutions were obtained, demineralized water was added to make stock solution of known molarity. The required volume of uranyl nitrate and lanthanum nitrate solutions were mixed as per the stoichiometry and citric acid (2 times the cationic molarity) was added as a chelating agent and fuel to the solution. The solution was slowly evaporated to form gel using a hot plate. Raising the temperature to around 573 K initiated the self ignition to form a fluffy precursor powder. It was then slowly heated in furnace up to 973 K to remove residual carbon. Later the powders were compressed to form pellets and heated at 1173 K for more than 30 hours in air atmosphere with intermittent grindings. The part of the compounds was heated at higher temperature up to 1573 K in air for further analysis.

### 2.2 Characterization

Few of the synthesized compounds in the series were analyzed for the absolute concentration of U and La using a high resolution ICP-OES (Ultima 2 Model, HORIBA Scientific (Jobin Yvon Technology), France). The analysis results are given in the supporting document (table ST1). The room temperature X-ray diffraction (XRD) patterns of the compounds were recorded using Rigaku SmartLab X-ray diffractometer using Cu K-alpha radiation. Thermal expansion and thermal phase transitions was studied by temperature dependent XRD using same XRD instrument with Rigaku HT1500 high temperature attachment. The temperature was kept iso-thermal while recording the XRD at required temperatures and interval ramp was 10 K/min. The chamber was continuously flushed with zero grade air with 100ml/min.

Crystal Structure refinement of the new phase was carried out using XRD. Phase quantification was carried out using Rietveld refinement using FULLPROF suite [18].

To investigate the thermal stability and oxidation behavior, Thermogravimetric analysis (TGA) was carried out using Mettler-Toledo TGA instrument. TGA was recorded in air up to 1273 K with 10 K/min ramp. For few of the samples, uranium to oxygen ratio (O/U) was calculated using UV-VIS spectrophotometric method and the oxygen to metal ratio (O/M) was calculated accordingly considering fixed oxidation state for La. Here the solid sample was dissolved in ortho-phosphoric acid (OPA) to retain the oxidation state of U in solution form. Later equal proportion of demineralized and de-oxygenated water was added to reduce the viscosity and hygroscopic nature of the concentrated OPA solution and equilibrated for few hours. The absorption spectra of the solutions were measured in the spectrophotometer in the visible range from 360 to 720 nm using Shimadzu UV-1700 PharmaSpec UV-visible spectrophotometer. The spectra were linearly fitted with the prior recorded and normalized spectra of U(IV) and U(VI). The coefficients of linear fitting are divided by specific absorptivity of U(IV) at 660nm and U(VI) at 420nm to find concentration of the U(IV) and U(VI) respectively and O/U is determined. As a reference, known standards like $KUO_3$ and $U_3O_8$ were also simultaneously analyzed along with the samples to verify the results. Further details of this method is given elsewhere [19].

**3. Results and Discussion:**

**3.1 XRD analysis of samples heated at 1173 K in air**

Fig. 1(a) shows the XRD pattern of the series of samples heated at 1173 K. The XRD pattern for y < 0.25 shows mixture of $U_3O_8$ and additional secondary phase with XRD lines similar to $U_3O_8$ (**henceforth called as new phase or as $M_3O_8$ to distinguish it from $U_3O_8$**). Fig. 1(b) shows that new phase starts emerging even at the lowest doping La-concentration of y=0.025. With increase in La- concentration, intensity of the $U_3O_8$ gradually decreases with increase in the intensity of new phases. At composition y=0.225, the $U_3O_8$ phase completely vanishes to form pure new phase ($M_3O_8$, M=0.875U+0.225La). To estimate the phase percentage of the two phases in these samples, phase quantification was done by Rietveld refinement of XRD data and equation (eq. 1) was used for calculating the weight percentage.

$$W_i = \frac{S_i Z_i M_i V_i / t_i}{\sum_i (S_i Z_i M_i V_i / t_i)} \quad -(1)$$

where, S is the refined scaling factor, Z is formula units per unit cell, M mass of the formula unit, V is the volume and t is the Brindley particle absorption contrast factor (neglected in the current calculation as absorption is similar for the phases considered).

As the new phase has the XRD pattern very similar to α- $U_3O_8$, the structure model of α- $U_3O_8$ with C2mm space group is used for the structure refinement of new phase where both the uranium positions are substituted with (0.875U+0.225La). This assumption is convincing as $U_3O_8$ reflections in the XRD pattern are below detection level at composition y=0.225. As the series representative, Fig. 2 shows the fitted XRD pattern for the composition y = 0.1. Similar quantitative phase analysis was done for rest of the compositions in the range y ≤ 0.225. Using the obtained weight percentages of the $U_3O_8$ and new phase (formula used for calculation $(U_{0.775}La_{0.225})_3O_8$), molar ratio is calculated for the two phases and subsequently the La/U ratio is calculated (Table. 1). Details of this calculation are given in the supplementary document. Fig. 3 shows the plot of the mole percentage of $U_3O_8$ and $(U_{0.775}La_{0.225})_3O_8$ phases up to composition y ≤ 0.225. It is observed that is calculated La/U ratios from the XRD phase analysis are close to La/U ratio taken in the starting composition and thus validates the analysis.

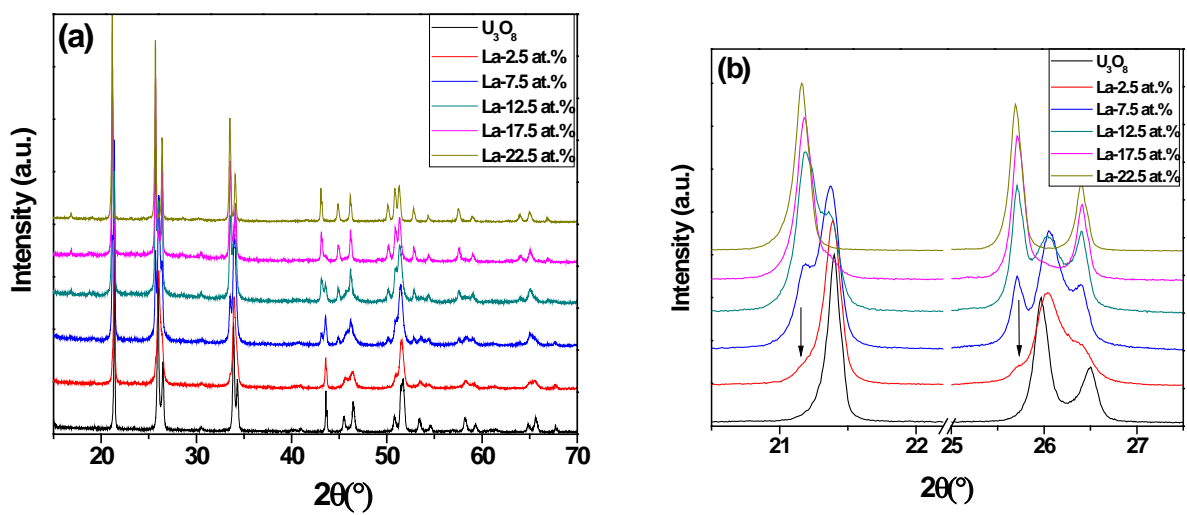

**Figure 1. (a)** XRD patterns for $U_{1-y}La_yO_{2+x}$ (y≤0.225) samples heated in air at 1173 K. Legend indicates the La- at% in the samples. **(b)** Magnified view of the first two XRD reflections. The arrow marks the inflection indicating the formation of secondary phase which gradually grows with the concentration of La- at%.

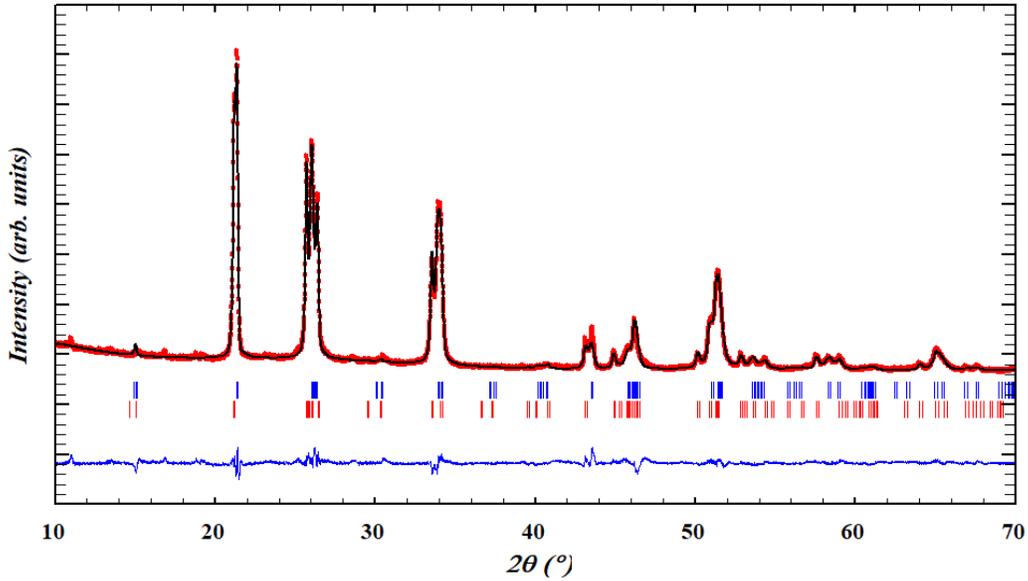

**Figure 2.** Fitted XRD pattern for the representative composition y=0.1 showing observed data (red scatter dot), calculated (black continuous line), difference curve (blue continuous line), upper tick mark in blue for $U_3O_8$ and lower tick marks in red for new phase.

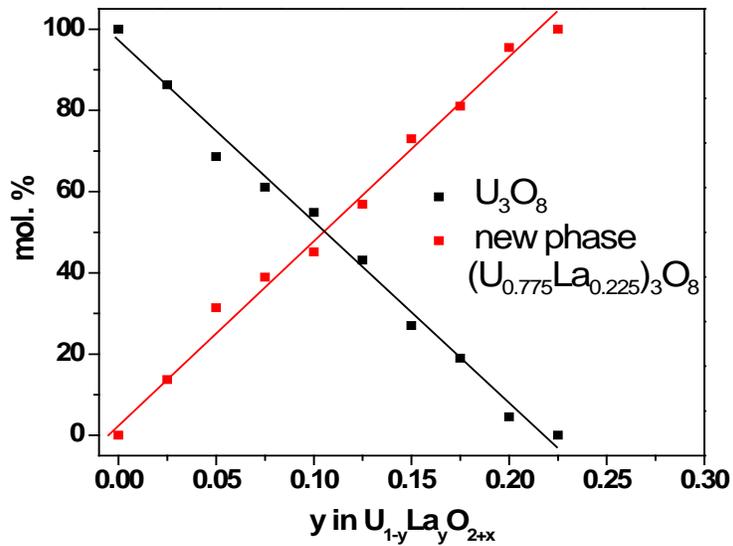

**Figure 3.** Mole percentage of $U_3O_8$ & $(U_{0.775}La_{0.225})_3O_8$ (for y ≤ 0.225) plotted against La- concentration in $U_{1-y}La_yO_{2+x}$ samples heated at 1173 K in air analyzed using phase quantification by Rietveld refinement of XRD data. Line is a guide for an eye.

**Table 1.** Phase quantification and cell parameters done using Rietveld refinement of XRD patterns for the $U_{1-y}La_yO_{2+x}$ samples heated at 1173 K in air. New phase $M_3O_8$ (M= 0.875 U+0.225 La); Fluorite phase ($La_{0.4}U_{0.6}O_{2+x}$)

| Sample $U_{1-y}La_yO_{2+x}$ | Phases present in mole fraction | Averaged La/U ratio calculated from XRD phase analysis (Original ratio taken) | Cell Volume of $U_3O_8$/ Cell parameter of Fluorite | Cell Volume of new phase $M_3O_8$ |
|---|---|---|---|---|
| y=0.025 | 0.86 $U_3O_8$ + 0.14 $M_3O_8$ | 3/97 (2.5/97.5) | 332.35(1) Å$^3$ | 342.53(8) Å$^3$ |
| y=0.05 | 0.69 $U_3O_8$ + 0.31 $M_3O_8$ | 7/93 (5/95) | 333.82(4) Å$^3$ | 342.50(4) Å$^3$ |
| y=0.075 | 0.61 $U_3O_8$ + 0.39 $M_3O_8$ | 8.7/91.3 (7.5/92.5) | 333.62(5) Å$^3$ | 342.36(5) Å$^3$ |
| y=0.1 | 0.55 $U_3O_8$ + 0.45 $M_3O_8$ | 10.2/89.8 (10/90) | 333.14(4) Å$^3$ | 342.15(5) Å$^3$ |
| y=0.125 | 0.43 $U_3O_8$ + 0.56 $M_3O_8$ | 12.8/87.2 (12.5/87.5) | 334.39(1) Å$^3$ | 342.35(8) Å$^3$ |
| y=0.15 | 0.27 $U_3O_8$ + 0.73 $M_3O_8$ | 16/84 (15/85) | 334.02(5) Å$^3$ | 342.29(7) Å$^3$ |
| y=0.175 | 0.19 $U_3O_8$ + 0.81 $M_3O_8$ | 18/82 (17.5/82.5) | 335.01(2) Å$^3$ | 342.12(4) Å$^3$ |
| y=0.2 | 0.05 $U_3O_8$ + 0.95 $M_3O_8$ | 21/79 (20/80) | 335.01(5) Å$^3$ | 342.13(2) Å$^3$ |
| y=0.225 | $M_3O_8$ | -- | -- | 342.39(1) Å$^3$ |
| y=0.25 | 0.69 $M_3O_8$ + 0.31 Fluorite | 26/74 (25/75) | 5.491(2) Å Fluorite | 342.18(2) Å$^3$ |
| y=0.275 | 0.57 $M_3O_8$ + 0.43 Fluorite | 28/72 (27.5/72.5) | 5.491(2) Å Fluorite | 342.28(2) Å$^3$ |
| y=0.3 | 0.42 $M_3O_8$ + 0.58 Fluorite | 30/70 (30/70) | 5.4911(2) Å Fluorite | 342.29(2) Å$^3$ |

The refined unit cell volume for the $U_3O_8$ and new phase $M_3O_8$ is given in the Table. 1. The volume of the new phase is slightly higher than the $U_3O_8$ and remains almost constant over all the compositions. This clearly shows that it is a line compound in U-La-O phase diagram and should have stoichiometry close to ($U_{0.775}La_{0.225})_3O_8$) and have unit cell volume is larger due to substitution of La. Additionally, it shows that there is negligible solubility of La in pure α- $U_3O_8$ phase up to 1173 K when heated in air. A structural analysis of the new phase is discussed ahead.

Fig. 4 shows XRD patterns for the compositions above y>0.225. It shows that above this La-concentration, there exists a two-phase mixture of the new phase ($U_{0.775}La_{0.225})_3O_8$) and the fluorite phase. With La-concentration the proportion of fluorite phase increases and from the Fig. 4 it is seen that even at end composition y=0.3, appreciable amount of new phase still remains in the mixture. Like before, two phase Rietveld refinement was carried out for XRD pattern for samples y=0.25, 0.275, 0.3. Roughly extrapolating the intensity ratio between the new phase and fluorite phase, it can be seen that at approximately y≈0.35 composition the intensity ratio converge to zero. So to start the Rietveld refinement, composition of the fluorite phase was taken as $La_{0.35}U_{0.65}O_2$. The weight fraction obtained from the preliminary refinement was then finally used to extrapolate the composition at which the new phases completely vanish and is close to y=0.4. In the final refinements composition of fluorite phase was fixed at $La_{0.4}U_{0.6}O_{2+x}$. Representative fitted XRD pattern for y= 0.275 is shown in the Fig. 5. Similar to the previous analysis, obtained weight percentage was used to calculate the mole percentage and La/U cation ratio was calculated for the mixture and is given in the Table 1. Table 1 data, clearly shows that the calculated and the initial sample composition match closely validating the molar concentration in bi-phasic region by Rietveld analysis.

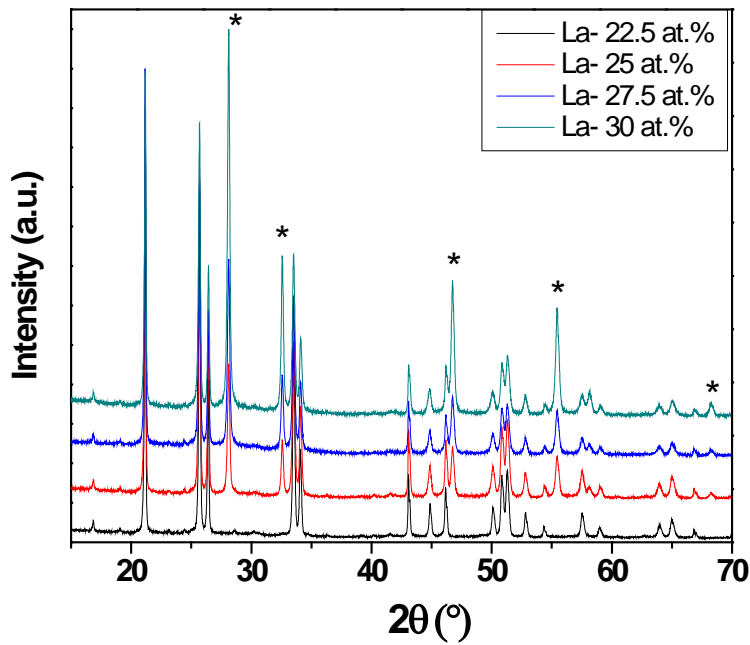

**Figure 4.** XRD patterns for $U_{1-y}La_yO_{2+x}$ (0.225 ≥ y ≥ 0.3) samples heated in air at 1173 K. Legend indicates the La- atom percentage in the samples. The '*' label indicate fluorite phase reflections.

The lattice parameters determined from the Rietveld refinement for y>0.225 compositions are given in the Table.1 and shows that the cell volume of the new phase remains identical to that obtained for y=0.225 and below confirming the retention of the pure new phase in studied conditions. The cell parameter of the fluorite phase obtained for y>0.225 compositions remains same at a=5.491(3)Å. This value is very close to the cell parameters of fluorite phase reported by D. C. Hill and F Hund et al. in the oxidizing condition for the composition close to y=0.4. Thus, our current phase analysis is in concurrence with the reported literature values [20-22].

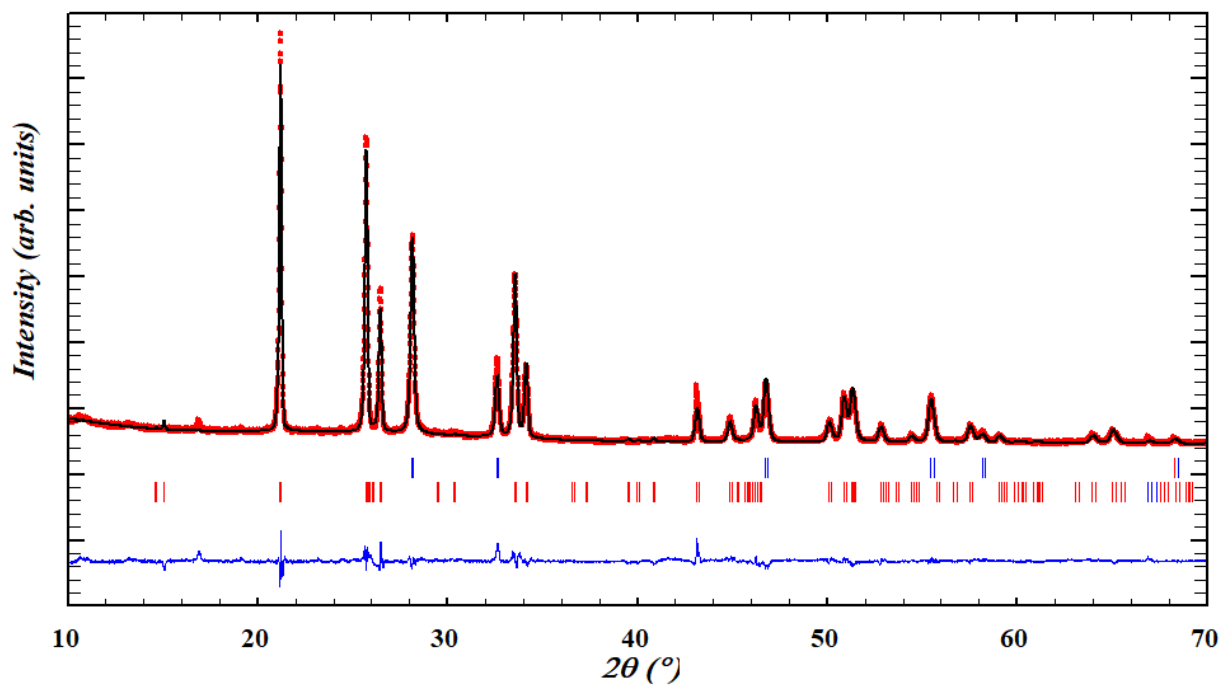

**Figure 5.** Fitted XRD pattern for the representative composition y=0.275 showing observed data (red scatter dot), calculated (black continuous line), difference curve (blue continuous line), upper tick mark in blue for Fluorite phase and lower tick marks in red for new phase.

To best of our knowledge, there is no literature report identifying a new line compound in U-La-O system which has XRD pattern very similar to α-$U_3O_8$. The reason for this may be the synthesis procedure or the synthesis temperatures adopted in the literature differ from the current adopted procedure. In general, there are two synthesis routes adopted in the literature for preparing $MO_{2+x}$ samples in lanthanide substituted urania phases, a) mixing the dry reagents in stoichiometry and heating the reactant mixture in reducing atmosphere to form fluorite phase $MO_{2-x}$ or $MO_{2.00}$ and later oxidizing the products to form $MO_{2+x}$ phases or b) mixing the dry reagents in stoichiometry and heating the reactant mixture in oxidizing atmosphere either in pure $O_2$ or dry air. In most reported cases the former procedure always leads to formation of $U_3O_8$ + fluorite phase mixtures or fluorite phase depending upon the initial composition, while the products formed in the later procedure depends upon the synthesis temperature. D. C. Hill observed formation of $U_3O_8$ below y=0.2 composition and fluorite phase for composition 0.2 < y < 0.8 for samples prepared at 1273 K in air [21]. Similar observation was done by F. Hund and U. Peetz where the samples heated in air at 1473 K showed fluorite phase formation in the region y=0.33 to 0.7 [22]. But in the same article, 100% fluorite phase was observed above y=0.4 as observed in current study. Oxidation of the vacuum reduced $U_{1-y}La_yO_{2\pm x}$ samples were done in air up to 1273 K using TGA instrument by H. Tagawa et al. [23]. The end products were mixture of $U_3O_8$ and fluorite up to y < 0.3; fluorite between 0.3 < y < 0.45. Zeynep Talip et al. heated the La-doped $UO_2$ samples in reduced atmosphere and subsequently heated them in the air at 500 K for 370h [24]. Due to low temperature oxidation they could partially achieve $M_4O_9$ type phase formation even at composition y=0.22 which is very close to our composition y=0.225, where we observe pure new phase formation with phase closely matching to $M_3O_8$ when heated at 1173 K. Bhupesh Kalekar et al. decomposed mixture of uranyl nitrate and lanthanum nitrate in TGA instruments up to 1073 K temperature in air [25]. For 20 mole% Lanthanum nitrate composition they observed formation of $U_3O_8$ and fluorite phase in the heated end product. But the carful observation of the XRD pattern given in the article (see Figure. 6 of the reference [25]), clearly indicates $U_3O_8$ peaks marked for 20 mole% Lanthanum nitrate composition but are broad compared to pure $U_3O_8$. This observation have two reasons; mixture of new phase $M_3O_8$ similar to $U_3O_8$ and pure $U_3O_8$ in the end product, which broadens the peaks and incomplete reaction in TGA experimental conditions where thermal equilibrium is not achieved. In the recent articles by Potts et al., Ln- doped $U_3O_8$ was prepared by precipitate method using uranyl/Ln-nitrate solutions with concentrated Ammonia solution. The dried precursor was heated up to 973 K in air [26]. The end product XRD showed broad XRD reflections as compared to pure α- $U_3O_8$ phase and was matched with $P\bar{6}2m$ space group similar to meta-stable α'- $U_3O_8$ phase which appears above 623 K temperature [27]. The authors also indicate the presence of mixture of phases with orthorhombic α- $U_3O_8$ and hexagonal α'- $U_3O_8$ structures in Ln-doped $U_3O_8$ samples to justify the asymmetric XRD peaks. This observation also indicates the mixture would be of the new phase $(U_{0.775}La_{0.225})_3O_8$ along with the remnant pure orthorhombic α- $U_3O_8$, which is also clearly evident in our samples as well. This shows that previous studies have missed the identification of the new line compound in the U-La-O system.

It has also been reported that $U^{6+}$ exhibits high vapor pressure in oxide forms, and care should be taken when heating uranium oxide samples in air at high temperatures [28, 29]. It has been observed that due to uranium loss at elevated temperatures, the stoichiometry may change and result in a systematic negative bias in U concentration measurements. This point is pertinent while considering the literature data where uranium oxides are treated at elevated temperatures above 1273K for long duration.

### 3.2 XRD analysis of samples heated at 1523 K in air

The samples heated at 1173 K were further heated at 1523 K in air for 10h. It is observed that after heating the samples at 1523 K in air, it forms mixture of $U_3O_8$ and fluorite phase in the region $0.025 \geq y \geq 0.25$, while for the other two compositions y=0.275 and y=0.3, pure fluorite phase formation occurs. This indicates that during the heating process at 1523 K, the new phase decomposes to $U_3O_8$ and fluorite phase which is also observed by thermogarvimetric studies and discussed in details ahead. The lattice parameter of the fluorite phase formed after treating at 1523 K remains almost similar and is equal to $a_{fluorite}$ =5.467(1). This value closely match with the lattice parameter reported by E. Stadlbauer et.al for the composition $U_{0.75}La_{0.25}O_{2.23}$ and by Tagawa et.al for the composition $U_{0.73}La_{0.27}O_{2.26}$ [23, 30]. The section of the XRD patterns of these samples is shown in the Fig. 6(a) for $0 \geq y \geq 0.25$. For the comparison purpose the XRD patterns are normalized with respect to the second highest reflection of $U_3O_8$ at $2\theta \approx 26°$. The figure clearly shows that the proportion of the fluorite phase gradually increases up to y=0.25 composition and above which fluorite phase is formed. Phase quantification by Rietveld analysis of XRD patterns was done for 1523 K heated samples. Table 2 shows phase analysis for 1523 K heated samples. Fig. 6(b) clearly shows decrease of mole% of $U_3O_8$ in the phase mixture with increase in Fluorite phase mole%. This shows that the lower limit of solubility of La in fluorite phase is y≈0.275 when heated in air above 1273 K and is similar to that reported by Tagawa et.al. [23]. Close analysis of the XRD pattern (see supporting document Fig. SF1) indicates the fluorite phase observed at composition $U_{0.725}La_{0.275}O_{2+x}$ belongs to $M_4O_{9-\delta}$ type phase with many weak supper-lattice reflections observed at lower angle. This is because the O/M in this composition reaches value close to 2.25 and the arrangement of additional oxygen atoms plays a dominant role in deciding the structure. It was observed by E. Stadlbauer et.al that all the compositions with O/M ≈ 2.23 show such super-lattice reflections similar to $M_4O_{9-\delta}$ type phase [23, 30]. Also, the XRD pattern for the composition y= 0.275 and 0.3, shows weak reflections in the lower angle along with the broad peaks except for $(2n, 0, 0)_{fluorite}$ reflections. This closely resembles to the XRD pattern generated for the $\alpha$-$U_4O_9$ structure reported by L. Desgranges et.al, where all the reflections, except $(2n, 0, 0)_{fluorite}$ are doublets [31]. Formation of $M_4O_{9-\delta}$ type phase was also validated by observation of $M_4O_9$ Raman band in 22 atom% La doped $UO_2$ sample by Zeynep Tali et.al [24].

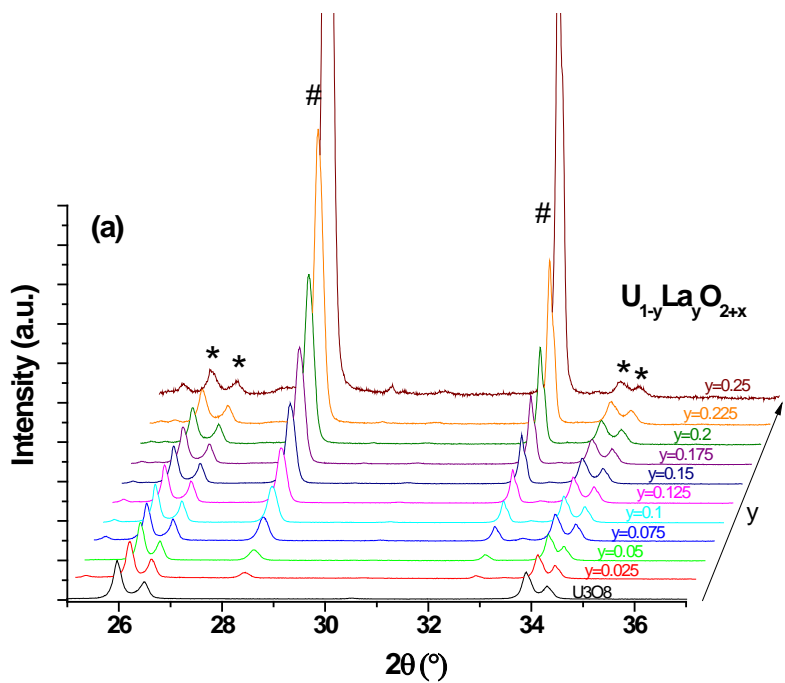

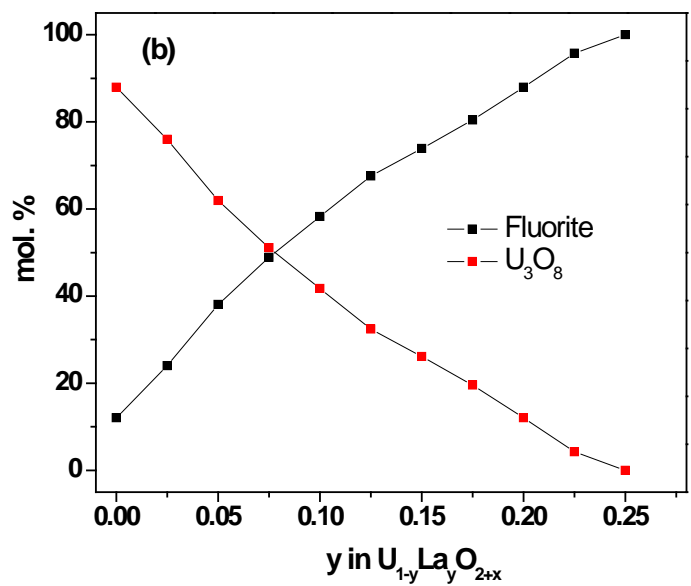

**Figure 6. (a)** Section of the XRD patterns for $U_{1-y}La_yO_{2+x}$ (0≥ y ≥ 0.25) samples heated in air at 1523 K and normalized with respect to the $U_3O_8$ reflection at 2θ≈26°. The patterns are stacked with increasing La concentration. The fluorite and $U_3O_8$ reflection are marked with * and # symbols respectively. **(b)** Mole percentage of $U_3O_8$ and fluorite phase plotted against La- concentration in $U_{1-y}La_yO_{2+x}$ samples heated at 1523 K in air analyzed by using phase quantification by Rietveld refinement of XRD data.

**Table 2.** Phase analysis done using Rietveld refinement of XRD patterns for the $U_{1-y}La_yO_{2+x}$ samples for heated at 1523 K in air. Fluorite ($U_{0.75}La_{0.25}O_{2+x}$)

| Sample $U_{1-y}La_yO_{2+x}$ | Phases present in mole fraction | Averaged La/U ratio calculated from XRD phase analysis (Original ratio taken) |
|---|---|---|
| y=0.025 | 0.12 Fluorite + 0.88 $U_3O_8$ | 2.3/97.7 (2.5/97.5) |
| y=0.05 | 0.24 Fluorite + 0.76 $U_3O_8$ | 4.7/95.3 (5/95) |
| y=0.075 | 0.38 Fluorite + 0.62 $U_3O_8$ | 7.9/92.1 (7.5/92.5) |
| y=0.1 | 0.49 Fluorite + 0.51 $U_3O_8$ | 10.5/89.5 (10/90) |
| y=0.125 | 0.58 Fluorite + 0.42 $U_3O_8$ | 13.0/87.0 (12.5/87.5) |
| y=0.15 | 0.68 Fluorite + 0.32 $U_3O_8$ | 15.7/84.3 (15/85) |
| y=0.175 | 0.74 Fluorite + 0.26 $U_3O_8$ | 17.6/82.4 (17.5/82.5) |
| y=0.2 | 0.80 Fluorite + 0.2 $U_3O_8$ | 19.8/80.2 (20/80) |
| y=0.225 | 0.88 Fluorite + 0.12 $U_3O_8$ | 22.4/77.6 (22.5/77.5) |
| y=0.25 | 0.96 Fluorite + 0.04 $U_3O_8$ | 25.3/74.7 (25/75) |
| y=0.275 | Fluorite | -- |

### 3.3 Oxygen to Uranium ratio determination

Oxygen to Uranium ratio (O/U) of the new $M_3O_8$ phase was determined using absorption spectrophotometry. Three aliquots were dissolved for the analysis. Fig. 7(a) shows the observed and fitted absorption spectrum for the new $M_3O_8$ phase. The spectrum was fitted using the equation (1):

$$\text{Spectrum}_{sample} = a \cdot \text{Spectrum}_{U(IV)} + b \cdot \text{Spectrum}_{U(VI)} + c/\lambda^4 \quad -(2)$$

where parameters a, b, and c are determined through least-square fitting using Python code. $\text{Spectrum}_{U(IV)}$ and $\text{Spectrum}_{U(VI)}$ corresponds to normalized absorption spectra for pure U(IV) and U(VI), respectively. The last term is fitted as a background term corresponding to Rayleigh scattering. Since normalized spectra are used for fitting, corresponding concentrations of U(IV) and U(VI) can be directly calculated using a, b coefficients and O/U can be determined using following equations:

$$\left. \begin{array}{l} C_{U(IV)} = a / \varepsilon_{U(IV)} \text{ (g/L)} \\ C_{U(VI)} = b / \varepsilon_{U(VI)} \text{ (g/L)} \\ O/U = 2 + C_{U(VI)} / (C_{U(IV)} + C_{U(VI)}) \end{array} \right\} \quad -(3)$$

The absorptivity constants (ε) are 0.1394(6) L.g$^{-1}$·cm$^{-1}$ for U(IV) at 660 nm and 0.04797(2) L.g$^{-1}$·cm$^{-1}$ for U(VI) at 420 nm wavelengths, respectively. The corresponding contributions from U(IV) and U(VI) normalized absorption spectra is shown in Fig. 7(b). The average O/U value determined for the new M$_3$O$_8$ phase is 2.89(2) and corresponds to the O/M = 2.89(2)*0.775 + 1.5* 0.225 = 2.58(2) with corresponding stoichiometry U$_{0.775}$La$_{0.225}$O$_{2.58(2)}$.

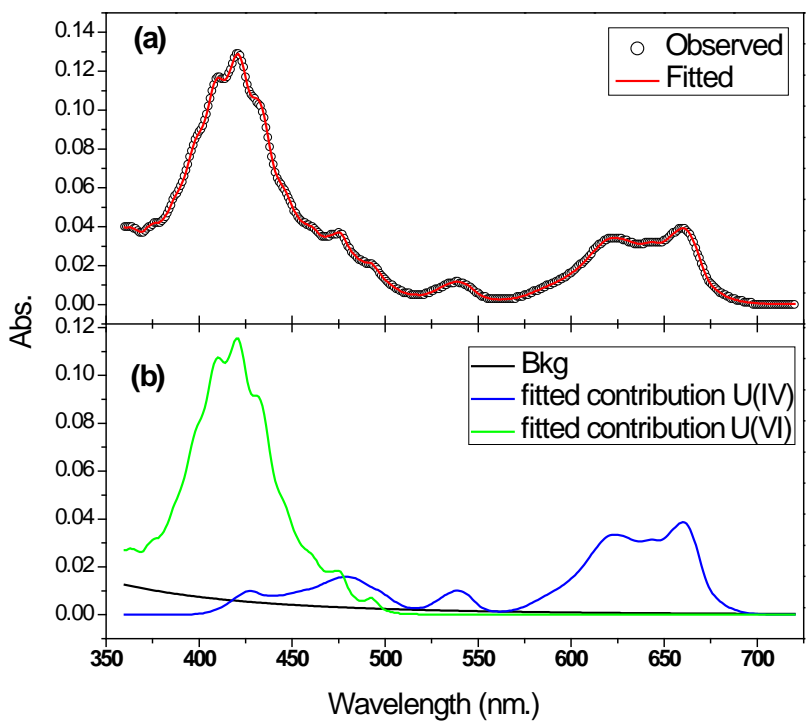

**Figure 7. (a)** Absorption spectrum recorded for the new phase with the composition U$_{0.775}$La$_{0.225}$O$_{2+x}$ and corresponding fit. **(b)** Fitted contribution from standard U(IV) and U(VI) absorption spectra.

### 3.4 Thermogravimetric analysis (TGA) of new phase

Two sets of TGA experiments were carried out on the composition U$_{0.775}$La$_{0.225}$O$_{2+x}$ heated at 1173 K which forms the new phase as discussed earlier. In the first TGA experiment, weighted quantity of sample was oxidized up to 1473 K in air and the thermogram is shown in Fig. 8(a). It is observed that the new phase decomposes above 1223 K with sharp weight loss. Thus, this new M$_3$O$_8$ phase is meta-stable and stable only below 1223 K. Percentage weight loss observed during the decomposition is 1.9 %. The XRD of the end product shows formation of fluorite + U$_3$O$_8$ phase mixture. Fluorite lattice parameters in this phase mixture match with the composition U$_{0.75}$La$_{0.25}$O$_{2+x}$ with x=0.22. The quantitative analysis of the phase mixtures were carried out using Rietveld refinement. The decomposition reaction is given below in reaction (4). O/M determined for the new $M_3O_8$ phase using absorption spectrophotmetry is 2.58(2), while for the oxidized end product is 2.27(2) which gives change in O/M of 0.32. The percentage weight loss calculated from change in O/M is 1.99 % which is close to observed in TGA i.e. 1.9 % during the decomposition.

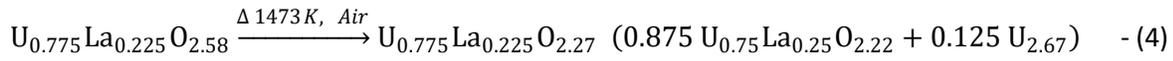

$$U_{0.775}La_{0.225}O_{2.58} \xrightarrow{\Delta\ 1473\,K,\ Air} U_{0.775}La_{0.225}O_{2.27}\ (0.875\ U_{0.75}La_{0.25}O_{2.22} + 0.125\ U_{2.67}) \quad - (4)$$

In the another TGA experiment new phase $U_{0.775}La_{0.225}O_{2.58(2)}$ was heated in the reducing atmosphere of Ar +8% $H_2$ at 1073 K for 1 h iso-thermal and is shown in Fig. 8(b). The sample shows weight loss due to loss of oxygen with 3.85% weight loss. The XRD of the end product shows formation of the pure fluorite phase. Considering the starting composition as $U_{0.775}La_{0.225}O_{2.58(2)}$ and the weight loss of 3.85% gives the final composition $U_{0.775}La_{0.225}O_{2-x}$ with x=0.03 i.e. $U_{0.775}La_{0.225}O_{1.97}$ and corresponding O/U=2.1 (see reaction 5). Thus in the current TGA experiment conditions, O/U lowers and saturates at 2.1 when heated at 1073 K for 1 h iso-thermal under reducing conditions. In our previous studies on RE- doped $UO_2$ samples, under reducing conditions sample forms hypo-stoichiometric fluorite phases and are highly unstable where uranium oxidizes in ambient room temperature conditions. Because of this high tendency of oxidation in hypo-stoichiometric Fluorite phase it acquires O/M close to 2.00 during storing conditions [5, 12].

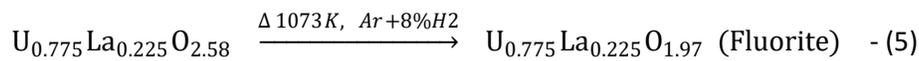

$$U_{0.775}La_{0.225}O_{2.58} \xrightarrow{\Delta\ 1073\,K,\ Ar+8\%H2} U_{0.775}La_{0.225}O_{1.97}\ (Fluorite) \quad - (5)$$

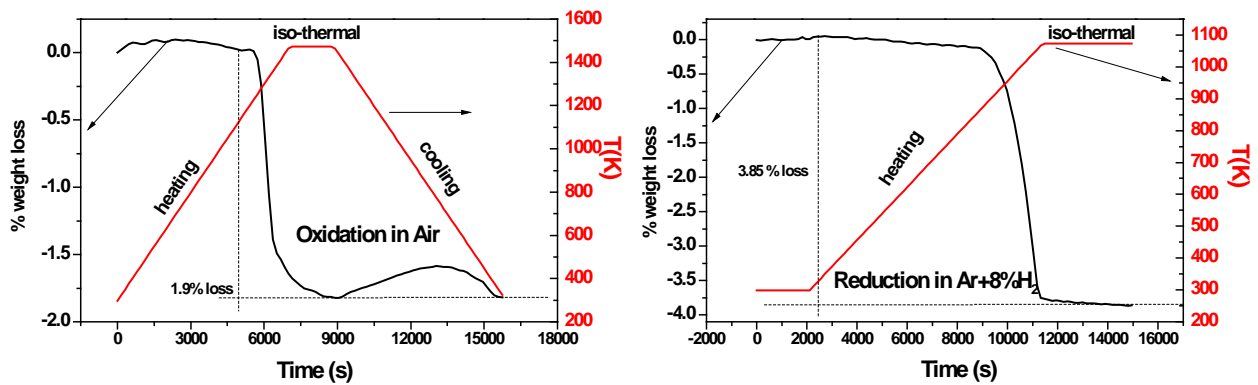

**Figure 8.** Thermogravimetric data for $U_{0.775}La_{0.225}O_{2.58(2)}$, **(a)** oxidation, heated in air up to 1473K & **(b)** reduction, heated in Ar+8% $H_2$ up to 1073K.

### 3.5 High temperature XRD study

As a supplementary technique to TGA experiment, High temperature XRD (HT-XRD) measurement was carried out on new phase $U_{0.775}La_{0.225}O_{2.58(2)}$. Sample was loaded on Pt sample holder plate and in-situ

XRD were recorded during heating from 300K to 1473K with the interval steps of 100K in air. Fig. 9 shows the XRD patters recorded at different temperatures. It is clearly seen that the XRD pattern of new phase $U_{0.775}La_{0.225}O_{2.58(2)}$ remains the same up to 1173K with no evidence of fluorite phase formation. Above this temperature, fluorite phase formation starts and gradually increases with increase in temperature. It is also evident from the XRD analysis that above decomposition temperature, the cell volume of the $U_{0.775}La_{0.225}O_{2.58(2)}$ phase suddenly drops and corresponds to $U_3O_8$. This observation is in agreement with the TGA analysis and shows that $U_{0.775}La_{0.225}O_{2.58(2)}$ phase is a metastable phase having mixed uranium valence.

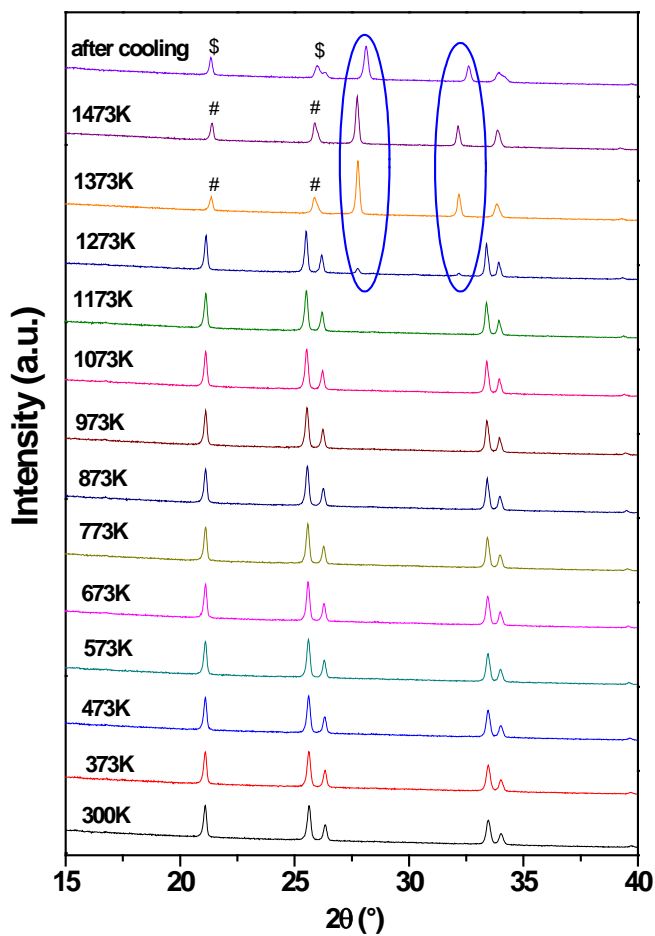

**Figure 9.** HT-XRD patterns for $U_{0.775}La_{0.225}O_{2.58(2)}$ recorded in air indicating compound decomposition above 1173K to form fluorite phase (highlighted in blue regions) and $U_3O_8$ (# -high temperature polymorph, $ -room temperature polymorph)

### 3.6 Crystal Structure analysis of new phase $U_{0.775}La_{0.225}O_{2.58(2)}$

As discussed above in the section 3.1, composition close to $U_{0.775}La_{0.225}$ shows formation of the new phase. All the high intensity reflections in the XRD pattern are very similar to that of α-$U_3O_8$ phase except very weak intensity super-lattice reflections. This may be the reason that many of the previous reports investigating U-La-O system have missed or ignored the weak reflections and interpreted this phase as $U_3O_8$ type phase with solubility of lanthanum. In our current study, we thoroughly investigated these weak reflections in XRD and attempted to identify the crystal structure of this new meta-stable phase $U_{0.775}La_{0.225}O_{2.58(2)}$. The preliminary indexing of the XRD pattern was done using α-$U_3O_8$ crystal model and obtained cell parameters are (a = 6.74(3) Å, b= 12.11(3) Å, c = 4.19(3) Å, V = 342.3(1) Å$^3$). The higher cell volume of $U_{0.775}La_{0.225}O_{2.58(2)}$ compared to α-$U_3O_8$ may be attributed to incorporation of larger ionic size of La$^{3+}$. Closely inspecting the XRD pattern and matching it with the ICDD database, two compounds were identified to have good match; $Pb_3U_{11}O_{36}$ and $Sr_3U_{11}O_{36}$ [32]. Both these compounds have U(VI) oxidation state and shown to have crystal structure related to α-$U_3O_8$. In general, all these compounds can be categorized in the series with general formula $A_nB_{n-2}O_{5n-4}$ with n = even numbers where α-$U_3O_8$, $Nb_5U_5O_{26}$ and $Pb_3U_{11}O_{36}$ / $Sr_3U_{11}O_{36}$ corresponds to n=4, n=6 and n=8, respectively. In α-$U_3O_8$ structure, both crystallographic sites for A and B cations are occupied by U atoms whereas in $Pb_3U_{11}O_{36}$ / $Sr_3U_{11}O_{36}$ the Pb/Sr and U occupy different sites due to charge and size difference to reduce the Medelung energy. Ordering between A and B cations and consequent oxygen anion ordering creates these higher super-structures. In this series, pentagonal bipyramids and/or octahedra are interconnected in layers and two such layers are connected to each other by vertices to form 3-dimentional structures. Fig. 10 shows the comparison between α-$U_3O_8$ and $Sr_3U_{11}O_{36}$ crystal structures. The similarity of repeating sets of seven polyhedra is marked with dotted line where α-$U_3O_8$ structure has all connected pentagonal bipyramids (marked as A) and connected to similar such pattern in the lower layer. In $Sr_3U_{11}O_{36}$ the connecting pattern of the polyhedra remains similar except the upper three pentagonal bipyramids are occupied by strontium atoms shown in green color and lower four pentagonal bipyramids are replaced by uranium octahedra shown in grey color (marked as A') and further connected to another set of 7 uranium polyhedra with two of the pentagonal bipyramids replaced by octahedra (marked as B') and these are connected to lower layer by screw symmetry.

Though the XRD pattern closely match with that of the reported XRD patterns of $Pb_3U_{11}O_{36}$ and $Sr_3U_{11}O_{36}$, the Rietveld refinement of $U_{0.775}La_{0.225}O_{2.58(2)}$ phase using above crystal structure model shows minor differences. Fig. 11 shows the fitted XRD pattern and the inset shows the enlarged part with weak reflections. It can been seen that although the overall XRD pattern is well fitted, few week reflections are not indexed correctly like very first reflection is clearly a doublet and not addressed with this model. Also, intensity mismatch is observed in few weak reflections. This indicates much more complex superstructure for $U_{0.775}La_{0.225}O_{2.58(2)}$ phase with the possibility of anti-site disordering between La and U cations at A and B sites. We collected neutron diffraction pattern and tried to refine the structure with oxygen atom positions but the refinement resulted in non-physical bond lengths for La-O bonds with $Pb_3U_{11}O_{36}$ and $Sr_3U_{11}O_{36}$ crystal structure models. Also, another issue with neutron diffraction refinement is, very low contrast between La and U coherent scattering cross-section and the issue of non-physical bond lengths for La-O bonds may be attributed to the same possibility of anti-site disordering between La and U cations at A and B sites. Due to this, the neutron diffraction study is excluded in the current discussion. The exact crystal structure of the $U_{0.775}La_{0.225}O_{2.58(2)}$ phase remains open and out of scope of current study.

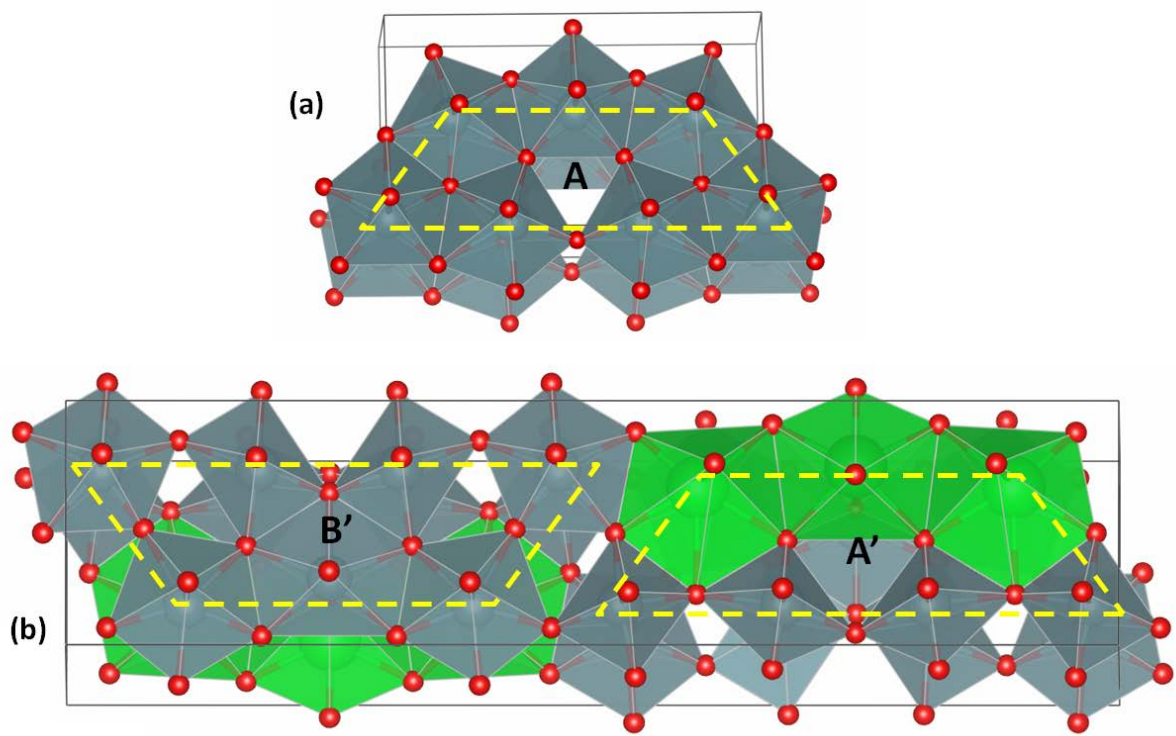

**Figure 10.** Crystal Structure of (a) α- $U_3O_8$ and (b) $Sr_3U_{11}O_{36}$. In α- $U_3O_8$, all the uranium has pentagonal bi-pyramid coordination shown in grey color while in $Sr_3U_{11}O_{36}$, uranium (grey) and strontium (green) exists in both pentagonal bi-pyramid and octahedral coordination. The repeating pattern of connecting polyhedrons is shown in yellow dash lines and shows the resemblance in both crystal structures.

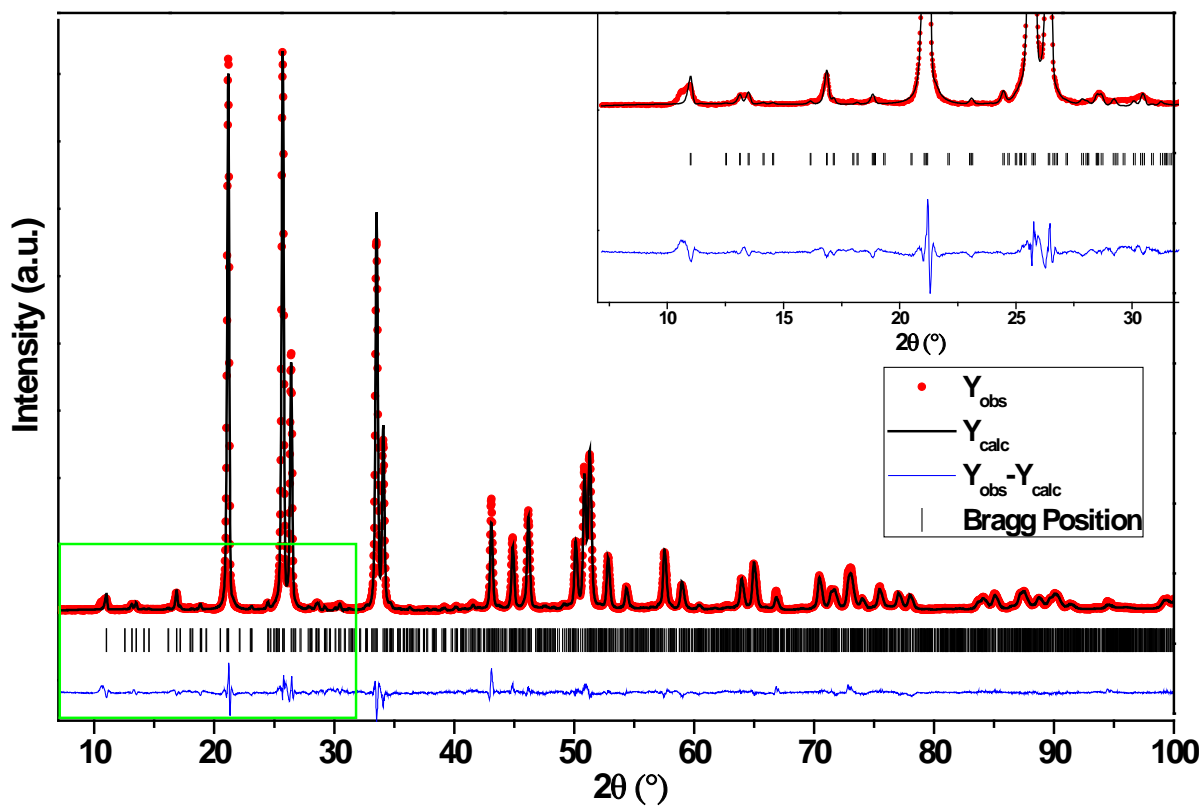

**Figure 11.** Fitted XRD pattern using Rietveld refinement for $U_{0.775}La_{0.225}O_{2.58(2)}$ phase using reported crystal structure model for $Sr_3U_{11}O_{36}$ compound. Inset shows the enlarged low angle region with weak super-lattice reflections.

### 3.7 Effect of ionic radii on $La_3U_{11}O_{36}$ crystal structure

In order to further investigate the cation size effect on the new phase with stoichiometry $La_3U_{11}O_{36}$, we attempted to synthesize this phase using other smaller size lanthanides like $Nd^{3+}$, $Sm^{3+}$, $Gd^{3+}$ and the smallest $Y^{3+}$ cation. Same synthesis route was adopted as used in U-La-O series given in experimental section. It is observed that these trivalent cations show formation of the new phase similar to $La^{3+}$ except for $Y^{3+}$. The respective XRD patterns for $Nd^{3+}$, $Sm^{3+}$, $Gd^{3+}$ are shown in the Fig. 12(a). Like $La_3U_{11}O_{36}$, $Nd_3U_{11}O_{36}$ phase was formed with same heating conditions, but it took much more heating cycles with intermittent grinding steps for Sm and Gd. In particular, it is evident from XRD pattern of Gd-sample that even after many heating cycles, pure $Gd_3U_{11}O_{36}$ phase conversion is incomplete. While for $Y^{3+}$ there was no formation of $Ln_3U_{11}O_{36}$ phase and shows formation of Fluorite phase with mixture of $U_3O_8$. Thus it is observed that the tendency to form $Ln_3U_{11}O_{36}$ phase decreases with the decrease in $Ln^{3+}$ cation size. Current study indicates, size of $Gd^{3+}$ can be considered as the limit for the formation of

$Ln_3U_{11}O_{36}$ phase. Fig. 12(b) shows the cell volume of $Ln_3U_{11}O_{36}$ phase plotted with respect to ionic size of $Ln^{3+}$, following linearity according to Vegard's law.

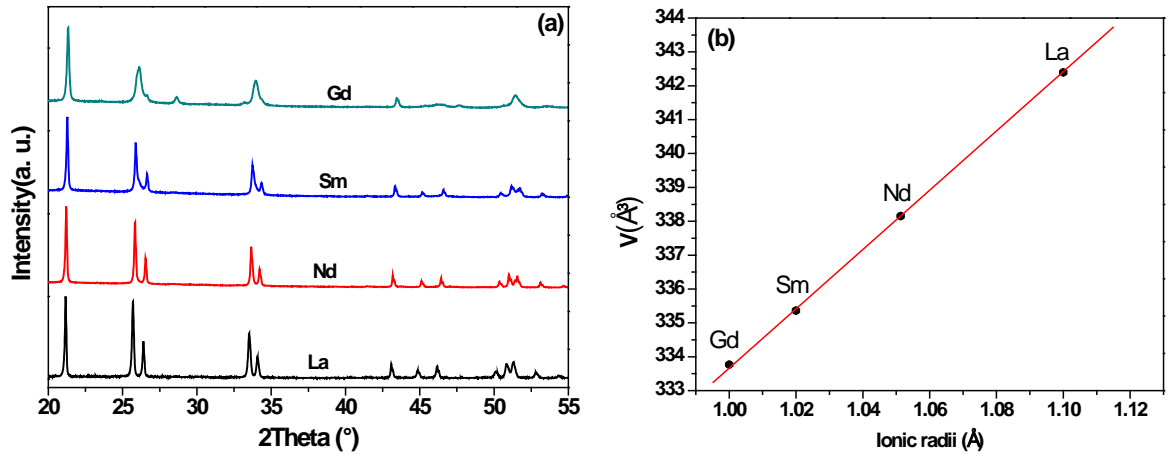

**Figure 12. (a)** XRD pattern for the new $Ln_3U_{11}O_{36}$ type phases (Ln = La, Nd, Sm, Gd). **(b)** Linear variation of Cell volume plotted against ionic radii of $Ln^{3+}$ in 7-coordination number.

## 4. Conclusion

Compositions $U_{1-y}La_yO_{2+x}$ (y=0.025, 0.05,…0.3) were synthesized in oxidizing environment and U-La-O system is re-investigated. The phase relations were established at and above 1173 K and XRD Rietveld analysis was used for the complete phase quantification within the studied composition range at 1173 K and 1523 K. The results show that there is no solubility of La in α-$U_3O_8$ phase even at lowest substitution composition, y=0.025. Within the studied composition range $U_{1-y}La_yO_{2+x}$ (y=0.025 to 0.3) heated at 1173 K in air, formation of novel phase with stoichiometry $La_3U_{11}O_{36}$ ($U_{0.786}La_{0.214}O_{2.57}$) was observed and reported for the first time. For all the compositions lower to y = 0.225, it forms mixture of $U_3O_8$ and $La_3U_{11}O_{36}$. Above y>0.225 to y=0.3, there is formation of $La_3U_{11}O_{36}$ and fluorite phase with composition $U_{0.6}La_{0.4}O_{2+x}$. The new phase identified as $La_3U_{11}O_{36}$ is a metastable phase with mixed uranium valence and decomposes when heated above 1223 K and forms $U_3O_8$ and fluorite phase ($U_{0.775}La_{0.225}O_{1.97}$) as end products. To investigate the phase relation above 1223 K, all the compositions were heated at 1523 K in air. In the region 0.025 ≥ y ≥ 0.25, it forms mixture of $U_3O_8$ and fluorite phases while for compositions y=0.275 and y=0.3, pure fluorite phase formation occurs. The new phase $La_3U_{11}O_{36}$ is a super structure of α-$U_3O_8$ and closely match to $Pb_3U_{11}O_{36}$ and $Sr_3U_{11}O_{36}$ compounds which fall in the series $A_nB_{n-2}O_{5n-4}$ with n = 8. Ionic size effect on the formation of this novel phase $Ln_3U_{11}O_{36}$ was studied for (Ln=Nd, Sm, Gd & Y). It shows that tendency to form this meta-stable phase decreases with decrease in ionic size of

Ln$^{3+}$ and ionic size of Gd being the lower limit to form Ln$_3$U$_{11}$O$_{36}$ phase while no such phase was observed for yttrium sample.

**Acknowledgement:**

The authors gratefully acknowledge Dr. Rajesh Pai, Head, Fuel Chemistry Division, BARC, for his support during this work. We also thank Miss. Leela Suthar, post-graduate student from Department of Chemistry, Jai Hind College, Mumbai, for her significant contribution to the synthesis of compounds described in this study during her post-graduate degree project. We also thank Dr. Abhijit Saha and Mr. Pranav Pathak from Radioanalytical Chemistry Division of BARC for ICP-OES analysis.